\documentclass{emulateapj}
\usepackage{apjfonts}

\slugcomment{}

\shorttitle{Transition from Young to Middle-aged SNR; N132D}
\shortauthors{Bamba et al.}

\begin{document}

\title{Transition from Young to Middle-aged Supernova Remnants:
Thermal and Nonthermal aspects of the SNR N132D}

\author{Aya Bamba\altaffilmark{1,2},
Yutaka Ohira\altaffilmark{3},
Ryo Yamazaki\altaffilmark{3},
Makoto Sawada\altaffilmark{3,4,5},
Yukikatsu Terada\altaffilmark{6},
Katsuji Koyama\altaffilmark{7,8},
Eric D. Miller\altaffilmark{9},
Hiroya Yamaguchi\altaffilmark{10,11},
Satoru Katsuda\altaffilmark{12,6},
Masayoshi Nobukawa\altaffilmark{13},
Kumiko K. Nobukawa\altaffilmark{14}
}
\altaffiltext{1}{%
Department of Physics, Graduate School of Science,
The University of Tokyo, 7-3-1 Hongo, Bunkyo-ku, Tokyo 113-0033, Japan
}
\altaffiltext{2}{%
Research Center for the Early Universe, School of Science,
The University of Tokyo, 7-3-1 Hongo, Bunkyo-ku, Tokyo 113-0033, Japan
}
\altaffiltext{3}{%
Department of Physics and Mathematics, Aoyama Gakuin University
5-10-1 Fuchinobe Chuo-ku, Sagamihara,
Kanagawa 252-5258, Japan
}
\altaffiltext{4}{%
X-ray Astrophysics Laboratory, NASA Goddard Space Flight Center,
Greenbelt, MD 20771, USA
}
\altaffiltext{5}{%
Department of Physics, University of Maryland Baltimore County,
1000 Hilltop Circle, Baltimore, MD 21250, USA
}
\altaffiltext{6}{%
Department of Physics, Science, Saitama University, Sakura, Saitama 338-8570,
Japan
}
\altaffiltext{7}{%
Department of Physics, Graduate School of Science, Kyoto University,
Kitashirakawa-oiwake-cho, Sakyo-ku, Kyoto 606-8502,
Japan
}
\altaffiltext{8}{%
Department of Earth and Space Science, Graduate School of Science, Osaka University,
1-1 Machikaneyama-cho, Toyonaka, Osaka 560-0043,
Japan
}
\altaffiltext{9}{%
Kavli Institute for Astrophysics \& Space Research, Massachusetts Institute of Technology, 77 Massachusetts Avenue, Cambridge, MA 02139, USA}
\altaffiltext{10}{%
NASA Goddard Space Flight Center, Code 662, Greenbelt, MD 20771, USA}
\altaffiltext{11}{%
Department of Astronomy, University of Maryland, College Park, MD 20742, USA}
\altaffiltext{12}{%
Department of Physics, Faculty of Science \& Engineering, Chuo University, 1-13-27 Kasuga, Bunkyo, Tokyo 112-8551, Japan}
\altaffiltext{13}{%
(Department of Teacher Training and School Education, Nara University of Education, Takabatake-cho, Nara, 640-8528, Japan}
\altaffiltext{14}{%
Department of Physics, Nara Women's University, Kitauoyanishimachi, Nara 630-8506, Japan}

\begin{abstract}
Supernova remnants (SNRs) are the primary candidate
of Galactic cosmic-ray accelerators.
It is still an open issue when and how
young SNRs, which typically exhibit strong synchrotron X-rays and GeV and TeV gamma-rays, undergo
the state transition to middle-aged SNRs
dominated by thermal X-rays and GeV gamma-rays.
The SNR N132D in the Large Magellanic Cloud
is an ideal target to study such a transition,
exhibiting bright X-rays and gamma-rays,
and with the expected age of $\sim$2500~yrs.
In this paper we present results of 
{\it NuSTAR} and {\it Suzaku} spectroscopy.
We reveal that N132D has a nearly equilibrium plasma with a temperature of $>$5~keV
or a recombining plasma with a lower temperature ($\sim$1.5~keV) and a recombining timescale 
($n_e t$) of 8.8 (7.0--10.0)$\times 10^{12}$~cm$^{-3}$s.
Together with 
the center filled morphology observed in the iron K line image,
our results suggest that N132D is now at transition stage
from a young SNR to middle-aged.
We have constrained the tight upper-limit of nonthermal X-rays.
Bright gamma-rays compared to faint nonthermal X-rays
suggest that the gamma-rays are hadronic in origin.
The spectral energy distribution from radio to gamma-rays 
shows a proton cut-off energy of $\sim$30~TeV.
These facts confirm that
N132D is in the transition from young to middle-aged SNR.
The large thermal energy of $>10^{51}$~erg and
accelerated proton energy of $\sim 10^{50}$~erg suggest
the supernova explosion might have been very energetic.
\end{abstract}


\section{Introduction}

Supernova remnants (SNRs) play crucial roles in physical processes at the universe:
e.g., they work as energy and heavy element suppliers,
and accelerate cosmic rays at their shocks.
They evolve in a rather short time scale in the interstellar medium.
Young SNRs expand with high shock velocity
and emit strong thermal and nonthermal X-rays from their shells.
Synchrotron X-rays are
excellent tools to detect accelerated electrons
\citep{koyama1995},
in addition to understand one of the key parameters, the magnetic field,
from its small-scale and sometimes time-variable structures \citep{vink2003,bamba2005,uchiyama2007},
and thus to resolve the origin of gamma-rays
\citep[][for example]{aharonian2007}.

On the other hand, middle-aged SNRs are fainter 
in both thermal and nonthermal emission,
and as a result, the study has developed rather recently.
{\it Fermi}/LAT has detected significant soft gamma-rays
from many old SNRs \citep{fermi2015},
implying that old SNRs still keep high energy particles into the space.
However, the detailed nature of these nonthermal emission is still poorly understood, 
since no synchrotron X-rays have been detected from middle-aged SNRs to date.
no synchrotron X-ray emission has been reported so far from middle-aged SNRs.

It is also noteworthy that most of the GeV-bright SNRs contain recombining plasma
\citep[e.g.,][]{yamaguchi2009},
implying that the plasma experienced rapid cooling during the SNR evolution.
Such a plasma emits strong radiative recombination continuum
and produces hard tails in the X-ray spectra.
Resolving this component will help to understand
the environment of particle acceleration sites.

N132D is the X-ray/gamma-ray brightest SNR in the Large Magellanic Cloud (LMC)
\citep{sutherland1995}.
It has a $\sim$14~pc ellipsoidal shell
expanding with the average speed of 1650~km~s$^{-1}$ \citep{morse1995}.
The SNR age is about 2500~yrs \citep{vogt2011}.
{\it Chandra} and XMM-{\it Newton} spectra showed
multi-temperature thermal emission
with strong emission lines from O to Fe
\citep{behar2001,borkowski2007}.
\citet{williams2006} reported 
a high dust density of $>$15~cm$^{-3}$,
a conclusion supported by the detection of 
a CO cloud with the mass of $\sim 2\times 10^{5}~M_\odot$
in the south of the SNR \citep{banas1997,sano2015}.
The southern part shows a enhanced thermal X-ray emission with circular shape, 
which supports the presence of dense ambient medium,
whereas the northern part has blown-out shape,
implying low-density interstellar medium in this direction.
Recently, GeV and Very high energy (VHE) gamma-rays
have been detected from N132D by {\it Fermi} and H.E.S.S.
\citep{fermi2015,hess2015}.
The gamma-ray spectrum is very hard like other young SNRs
such as Cas~A and RX~J1713$-$3946
\citep{abdo2010,abdo2011}.
Surprisingly, the measured luminosity in the 1--100~GeV band,
$\sim 10^{36}$~erg~s$^{-1}$ at 48~kpc \citep{macri2006},
is two orders of magnitude higher than
those of typical young SNRs.
Actually this is the highest luminosity among all known GeV SNRs
\citep{acero2016}.
Such a feature in the gamma-ray band implies that
N132D is an energetic cosmic-ray supplier.
Furthermore, it is difficult to make such bright gamma-rays
with leptonic models,
suggesting that gamma-rays are hadronic in origin.
Interestingly, we have no report of a synchrotron X-ray detection
from this source, probably because the soft X-ray spectrum is dominated  
by the luminous, high-temperature thermal X-rays up to 
Fe-K line band \citep{behar2001,borkowski2007,yamaguchi2014}.
We need sensitive observations in the hard X-ray band above 10~keV,
where thermal emission becomes much fainter.
{\it NuSTAR} is an ideal observatory for such a study
with its hard X-ray imaging system \citep{harrison2013}.
{\it Suzaku} also has a large effective area,
and low, stable background in the hard X-ray band, 
suitable for the nonthermal emission search 
\citep{mitsuda2007}.

In this paper, we present the first results of
the hard X-ray observation of N132D with {\it NuSTAR} and {\it Suzaku}.
\S\ref{sec:obs} and \S\ref{sec:results} describe 
the observation details and the analysis results, respectively.
In \S\ref{sec:discuss} we discuss the
condition of particle acceleration on N132D
together with radio, infrared, GeV and VHE gamma-ray results.
In this work, data reduction and analysis were performed with
HEADAS version 6.20 and XSPEC version 12.9.1.
For the spectral analysis,
atomic data base (ATOMDB) version 3.0.8
and the {\tt nei} version (NEIVERS) 3.0.7 were used.
Throughout the paper, we use 90\% error bars and confidence intervals.

\section{Observations and Data Reduction}
\label{sec:obs}

\subsection{{\it NuSTAR}}
N132D has been observed by the pixelated CdZnTe focal plane modules,
FPMA and FPMB, onboard {\it NuSTAR} with a single pointing
on 2015 Dec. 10--11.
The data were reprocessed with the calibration data base
(ver.20151008).
The cleaned events were extracted with the standard screening criteria,
except for the criteria for the passage of the South Atlantic Anomaly (SAA).
Since our target is faint in the hard band,
and the background count rate slightly elevated around the SAA,
we started the analysis with data filtered with
the parameters SAAMODE=STRICT and TENTACLE=yes.
The total exposure is 62.3~ks, which is 90\% of the nominal screening case.
The observation log is shown in Table~\ref{tab:obs}.

\subsection{{\it Suzaku}}

N132D has been observed by {\it Suzaku} several times
for calibration purposes.
We selected all data except for those taken in {\it Suzaku} initial operation.
The details of observations are shown in Table~\ref{tab:obs}.
{\it Suzaku} has several kinds of detectors,
four X-ray Imaging Spectrometers (XIS0--XIS3; \cite{koyama2007}) with
each at the focus of an X-Ray Telescope (XRT; \cite{serlemitsos2007}),
and a separate Hard X-ray Detector (HXD; \cite{takahashi2007}).
We concentrate on the analysis of XIS,
since {\it NuSTAR} has much better sensitivity than HXD.
We have checked the HXD data and found no significant signal.
Only three XISs operated in all of these observations.
XIS1 is a back-illuminated (BI) CCD,
whereas the others are front-illuminated (FI).
The XISs were operated in normal full-frame clocking mode
with spaced-row charge injection \citep{nakajima2008,uchiyama2009}.
The data was reprocessed with the
calibration database version 2016-02-14 for XIS
and 2011-06-30 for XRT.
We applied the standard screening criteria
to create the cleaned event list.
The total exposure is 240.3~ks.
The observation log is shown in Table~\ref{tab:obs}.

\section{Results}
\label{sec:results}

\subsection{{\it NuSTAR}}

\subsubsection{Images}

\begin{figure}
\epsscale{0.32}
\plotone{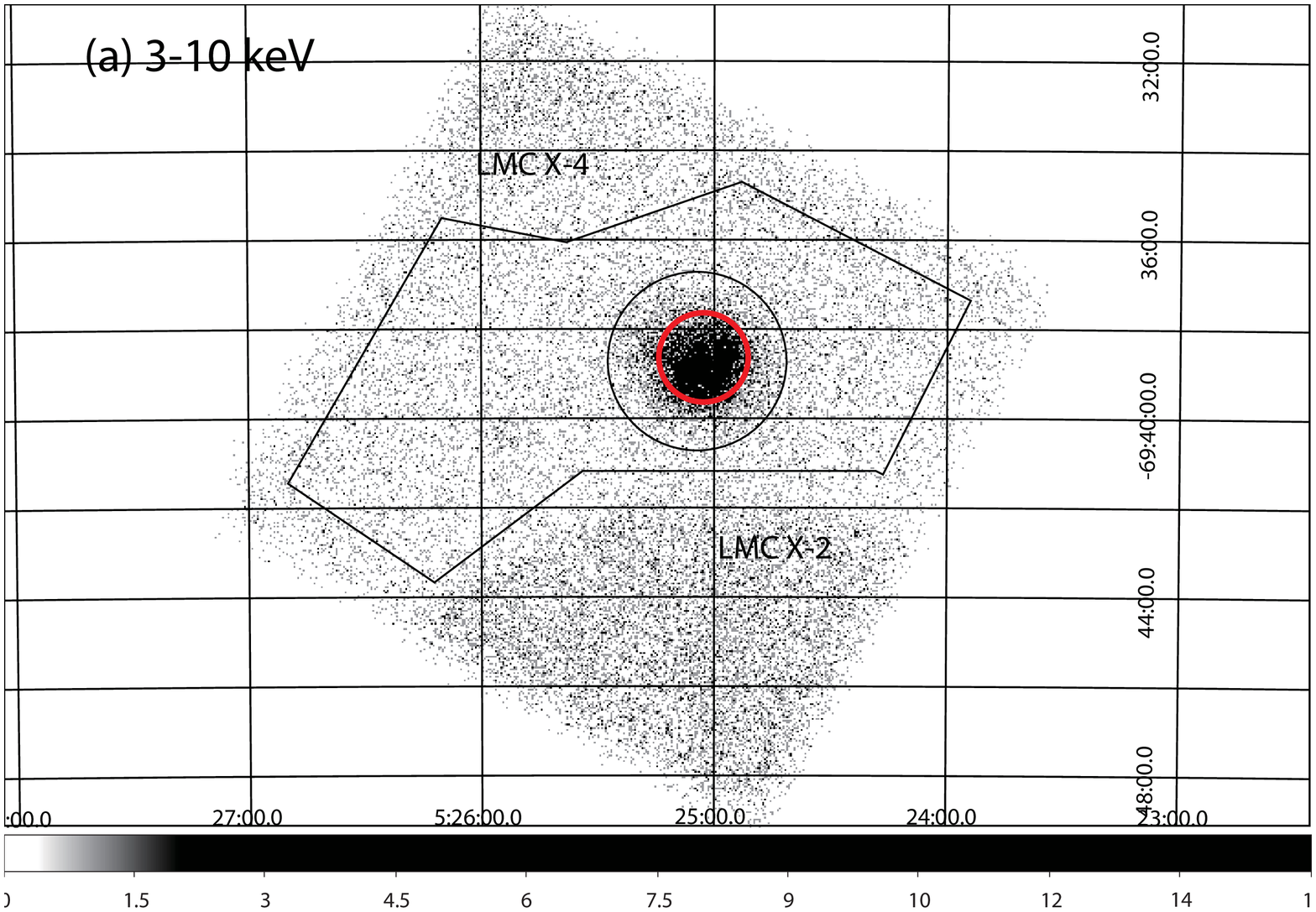}
\plotone{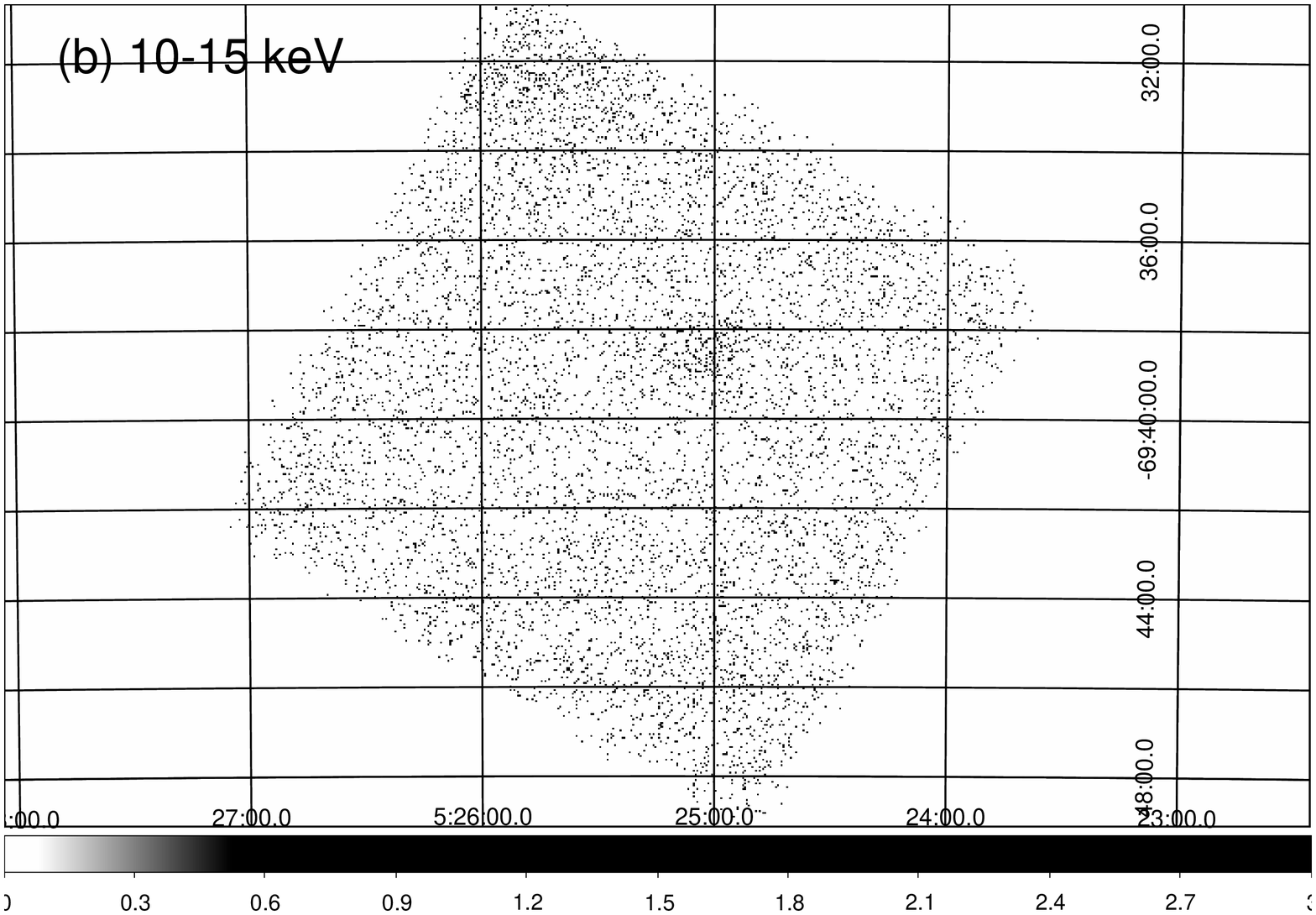}
\plotone{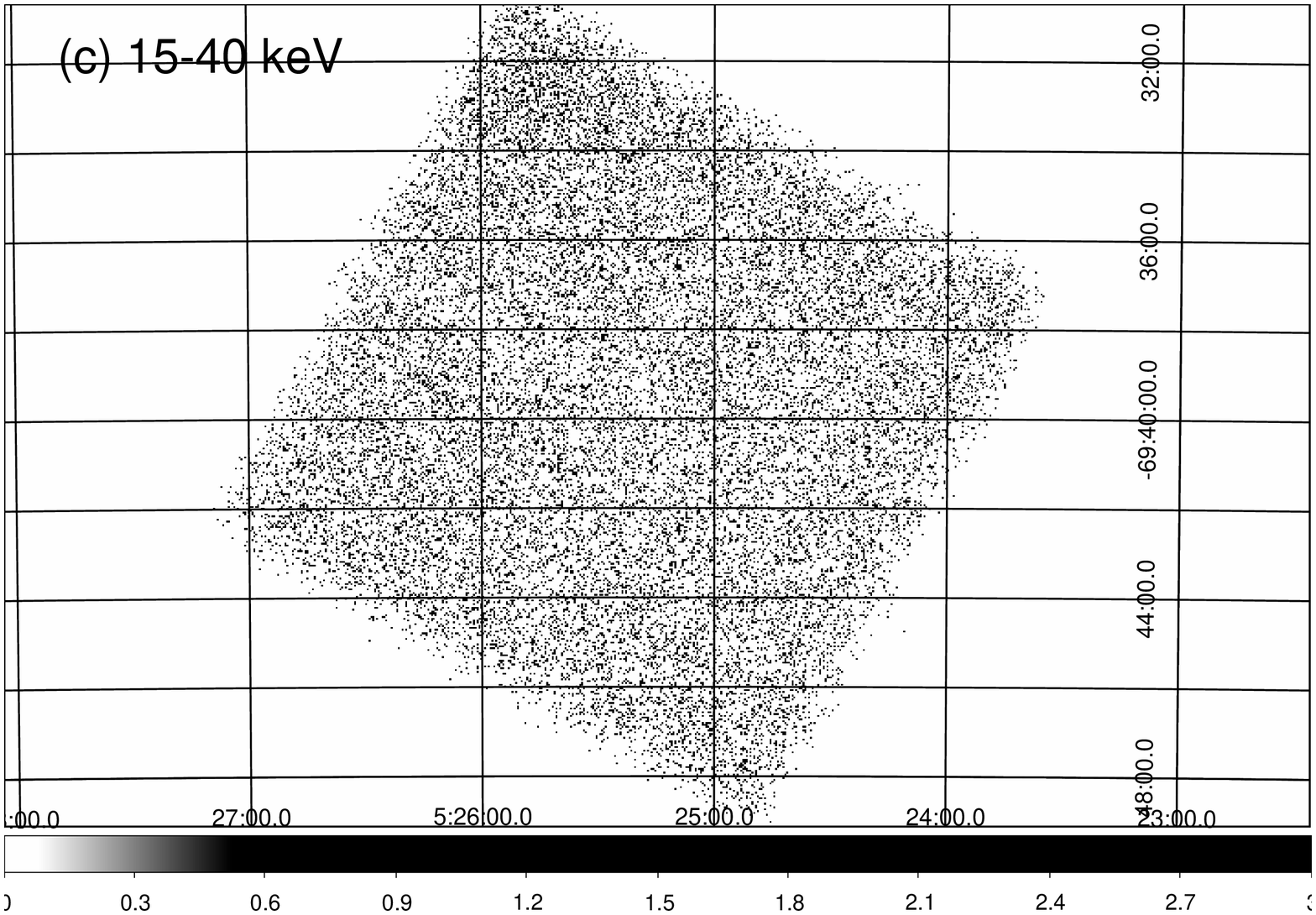}
\caption{%
The {\it NuSTAR} FPMA + FPMB images of N132D
in (a) 3--10~keV, (b) 10--15~keV, and (c) 15-40~keV band.
The images are in the linear scale and the coordinates are in J2000.
The stray lights from LMC~X-2 and LMC~X-4 are also marked.
Bold-red and thin-black regions represent source and background regions for
the spectral analysis.
}
\label{fig:nustarimages}
\end{figure}

Figure~\ref{fig:nustarimages} shows {\it NuSTAR}
3--10~keV (a), 10--15~keV (b), and 15--40~keV (c)
images,
where both FPMA and FPMB were used.
We can see clear enhancement on the N132D region below 15~keV,
whereas it is not clear above 15~keV.
The remnant has a size of $\sim$0.7$\times$ 1.0~arcmin$^2$,
which is smaller than the point spread function of {\it NuSTAR}
and we could not resolve the extension or structure.

We also found stray light from LMC~X-4 in both detectors
and that from LMC~X-2 in FPMB (see Figure~\ref{fig:nustarimages}).
Another nearby high mass X-ray binary, XMMU~J054134.7$-$682550,
can produce stray light
since it sometimes show the flaring \citep{liu2005,palmer2007}.
We have checked {\it Swift}/BAT observations \citep{krimm2013},
and found that it was in the steady state with the flux of only $\sim$1~mCrab,
and that it did not show any flare during our observation.
We thus ignore the stray light effect by XMMU~J054134.7$-$682550.

\subsubsection{Spectra}

The source photons were extracted from a 1.4~arcmin-radius region
centered on N132D,
whereas we selected the background region free
from other sources and stray lights,
as shown in Figure~\ref{fig:nustarimages}.
The response files are produced with the {\tt nuproducts} command
under the point source assumption.

\begin{figure}
\epsscale{0.7}
\plotone{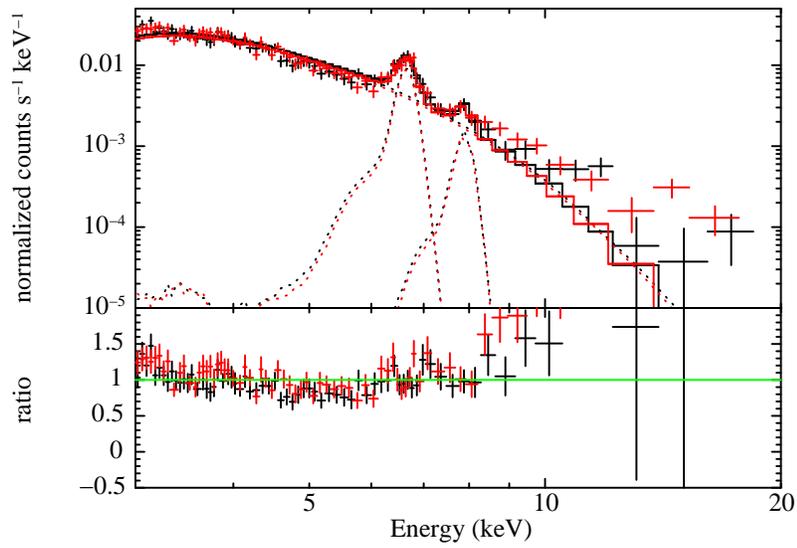}
\caption{Background-subtracted {\it NuSTAR} spectra.
Black and red crosses represent FPMA and FPMB data, respectively.
Solid and dotted lines show the best-fit model of total and each component.
The tails in the low energy side of lines are due to the detector response.
}
\label{fig:nustarspec}
\end{figure}

Figure~\ref{fig:nustarspec} shows the background-subtracted spectra of 
{\it NuSTAR} FPMA and FPMB.
One can see emission-like structures around 6.5 and 7.9~keV.
The former is reported as a K$\alpha$ emission line
from He-like Fe \citep{borkowski2007,yamaguchi2014},
whereas there is no report on the latter.
We fitted the spectra with a bremsstrahlung plus two Gaussian components.
The c-statistic \citep{cash1979} was used throughout this paper.
The background spectra were fitted simultaneously in XSPEC,
whereas background-subtracted spectra were used for the spectral plot\footnote{see https://heasarc.gsfc.nasa.gov/docs/xanadu/xspec/manual/node293.html}.
The best-fit models and parameters are shown in
Figure~\ref{fig:nustarspec} and  Table~\ref{tab:nustarlines}.
From the line center energies, we identified the 6.66~keV line
as He-like Fe K$\alpha$,
and 7.9~keV line as He-like Ni K$\alpha$ or He-like Fe K$\beta$.
One can also see the residuals above 10 keV,
which can be nonthermal hard tail or higher-temperature component.
The flux above 10~keV is 2.4 (2.0--2.8)$\times 10^{-14}$~erg~cm$^{-2}$s$^{-1}$.

\subsection{{\it Suzaku}}

\subsubsection{Images}

Figure~\ref{fig:suzakuimage} shows
the {\it Suzaku} XIS3 5--10~keV image.
We used only XIS3 since the part of XIS0 is not available 
and XIS1 has low efficiency and high background in this energy band.
One can see a point-like source in the center of the field of view.

\subsubsection{Spectra}

\begin{figure}
\epsscale{0.7}
\plotone{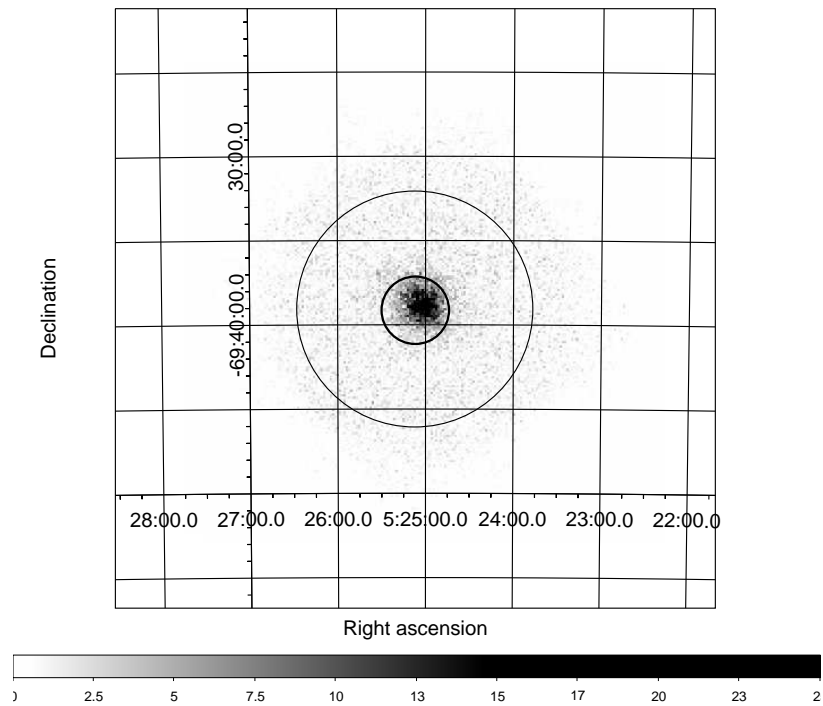}
\caption{%
XIS3 image of N132D in the 5--10~keV band in J2000 coodinates.
Neither vignetting correction nor NXB subtraction was performed.
Thick and thin circles shows the region for the source
and NXB estimation regions, respectively.
\label{fig:suzakuimage}}
\end{figure}

We extracted source photons from a circular region with a radius of 2~arcmin 
as shown in Figure~\ref{fig:suzakuimage}.
The background spectra mainly contain the Non X-ray Background (NXB) and
Cosmic X-ray Background (CXB) above 5~keV.
Other background components such as local hot bubble \citep{yoshino2009}
and LMC diffuse plasma \citep{points2001}
have a temperature of less than 1~keV, which is negligible in our band.
The NXB is generated by {\tt xisnxbgen} \citep{tawa2008}
for 7~arcmin radius region (see Figure~\ref{fig:suzakuimage})
and adjusted with the 11-14~keV count rate,
where we expect neither source nor CXB emission \citep{sekiya2016}.
We reproduced the CXB emission by the power-law model
with the photon index of 1.4
and the surface brightness in the 2--10~keV band of $5.4\times10^{-15}$~erg~s$^{-1}$~cm$^{-2}$~arcmin$^{-2}$ \citep{ueda1999}.
Since our target is point-like whereas the CXB is uniformly distributed,
we simulated the CXB spectrum with the assumption of uniform distribution
within 1.4~arcmin radius region
and fit it with the point source arf,
and used the best-fit parameters in N132D spectral fitting.

\begin{figure}
\epsscale{0.7}
\plotone{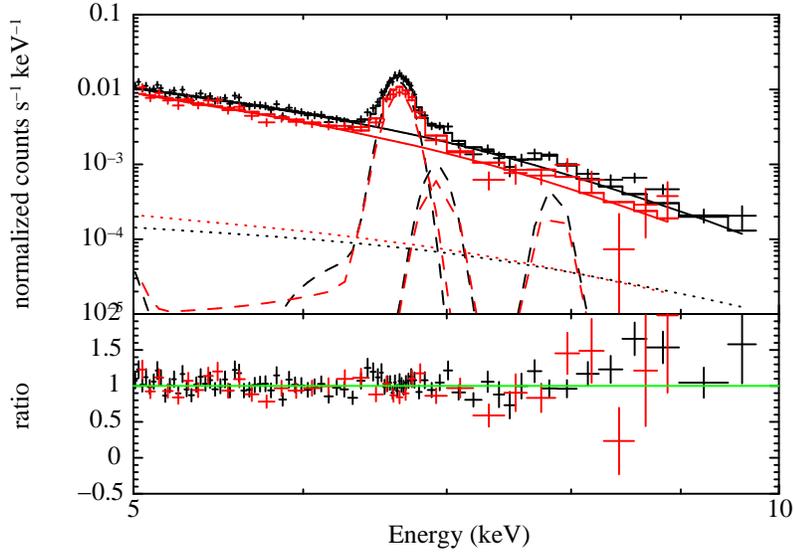}
\caption{NXB-subtracted {\it Suzaku} XIS spectrum.
Black and red crosses represent
FI (XIS0+XIS3) and BI (XIS1) data.
Solid, dashed, and dotted lines represent
bremsstrahlung, gaussian components, and the CXB model,
respectively.
The tails in the low energy side of lines are due to the detector response.
}
\label{fig:suzakuspec}
\end{figure}

Figure~\ref{fig:suzakuspec} shows the NXB-subtracted XIS spectrum above 5~keV.
One can see clear emission lines, similar to the {\it NuSTAR} results.
We fit the spectrum with bremsstrahlung plus 2 narrow gaussian model.
The fit returned cstat/d.o.f. of 167.5/111
with positive residuals around 6.9~keV,
which could not seen in the {\it NuSTAR} spectra
maybe due to the lack of statistics and energy resolution.
We added one more narrow gaussian to represent this structure,
which improved the cstat/d.o.f. to 145.4/109.
The line center energy agrees with the Fe Ly $\alpha$ emission line,
and this is the first detection of this line from N132D.
We still have small positive residuals in the high energy end,
which is consistent with the {\it NuSTAR} data.
Table~\ref{tab:suzakulines} shows the best-fit parameters.

\subsection{Combined spectral Analysis}

In the previous subsection, we reported the first detection of Fe Ly $\alpha$ emission 
from N132D, indicating that a substantial fraction of Fe atoms in this SNR is 
highly ionized. Keeping this in mind, we here constrain thermal parameters
with {\it NuSTAR} and {\it Suzaku} combined spectral analysis.
We start the fitting with 5--15~keV band.
We first introduce a {\tt vapec} model, a plasma in ionization equilibrium.
We used the photoionization cross-section table by \citet{balucinska1992}.
The abundances of Fe and Ni were treated as a common free parameter,
whereas those of the other elements were fixed to the solar values
\citep{anders1989}.
The best fit was obtained with an electron temperature of 2.31 (2.23--2.39)~keV and 
cstat/d.o.f. of 195.2/142,
leaving positive residuals around 6.9~keV due to the Fe Ly$\alpha$ line 
detected in the {\it Suzaku} spectrum.
This implies that the average charge balance is higher than the temperature
predicted by the ionization equilibrium model.
We thus added a second {\tt vapec} component  
and found that the fit was improved significantly (cstat/d.o.f. = 134.6/140),
with $kT$ of 1.1 (0.9--1.3) keV and 5.0 (4.5--6.3)~keV.
A higher charge balance is also achieved when the plasma is over-ionized.
We thus replaced the second vapec model with a vrnei model having a higher initial temperature to represent a recombining plasma.
This model also improved the fitting
with cstat/d.o.f. of 173.4/140.
We still have positive residuals in higher energy band,
which can be nonthermal emission from accelerated electrons.
We thus added a power-law component.
The photon index is fixed to 2.4, same to the gamma-ray emission.
The fitting was improved to cstat/d.o.f. of 155.8/139.

\begin{figure}
\epsscale{0.45}
\plotone{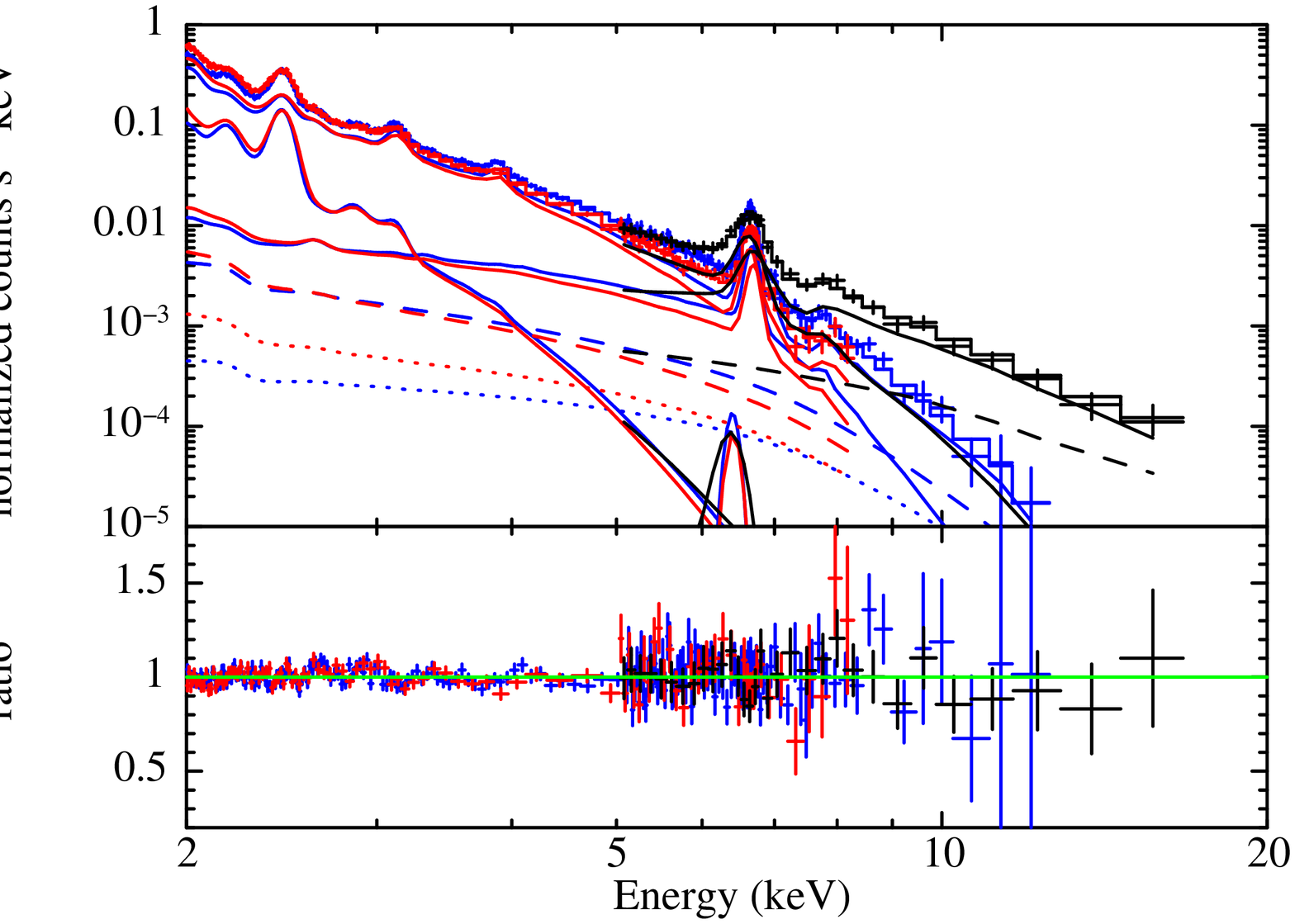}
\plotone{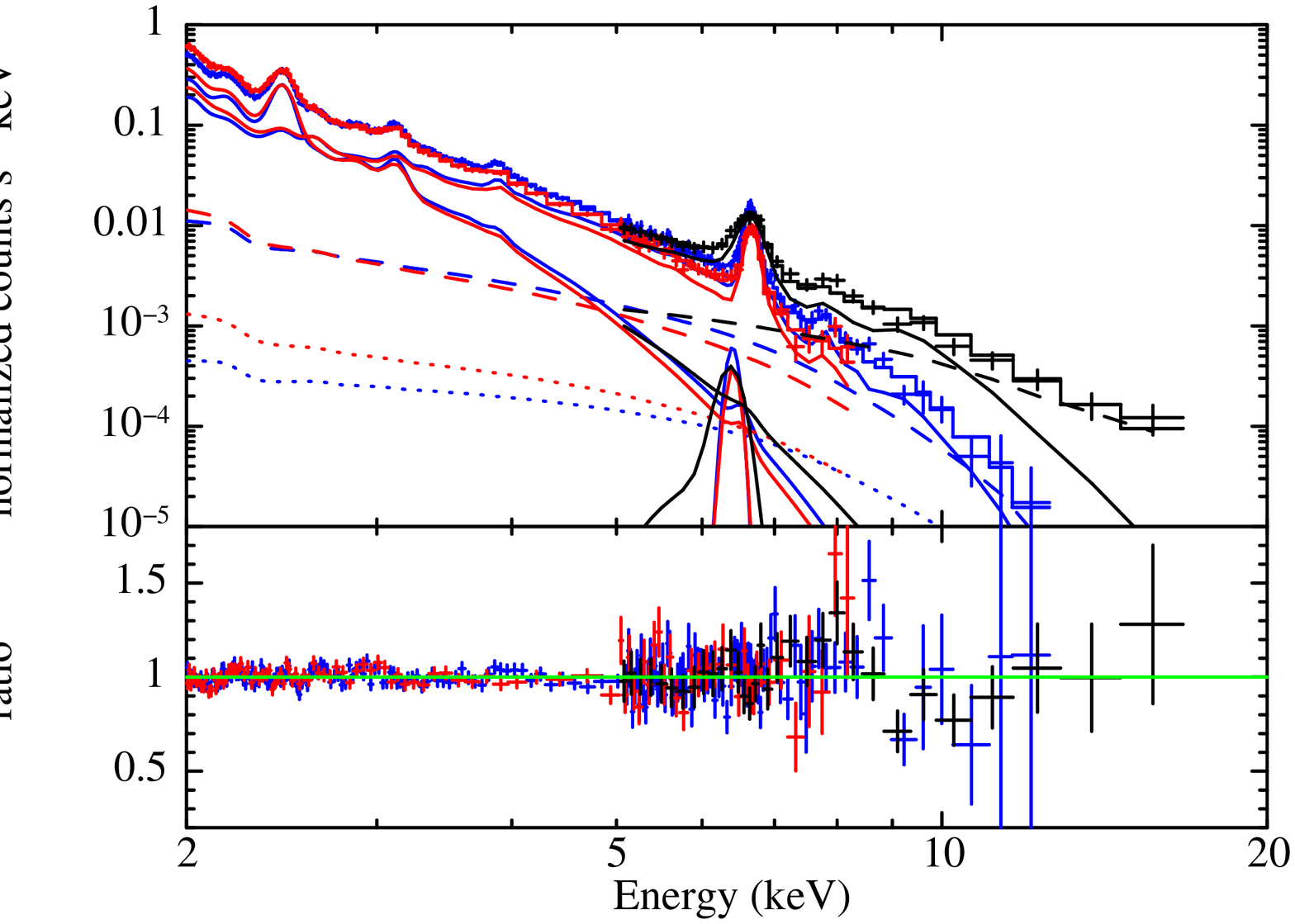}
\caption{%
Wide-band spectral fitting result for model (a) (left) and model (b)  (right) component models.
Black, red, and blue crosses represent
{\it NuSTAR}, {\it Suzaku} FI, and BI data set.
Solid, dashed, and dotted lines represents
thermal and power-law components, and the CXB for {\it Suzaku}.
\label{fig:wideband}}
\end{figure}

In the next step of adding the power law component, we extended
the fitting range to 2--15~keV band
to check the plasma condition of these higher temperature components.
In the energy band below $\sim$3~keV,
we already know that
there is at least one lower temperature component \citep{borkowski2007},
thus we added one more {\tt vapec} component 
with lower temperature.
Hereafter, we call ``model (a)" for the three {\tt apec} plasma model
and ``model (b)" for the two {\tt apec} plus a {\tt vrnei} plasma model.
We fixed the abundances of the low temperature compoet
to the LMC abundance \citep{russell1992},
but the best-fit model has significant residuals on Si and S lines.
We thus left the abundances of Si and S for the low temperature component
free.
The abundances of Si and S, and Fe and Ni of middle temperature component
were treated as common free parameters,
whereas the other elements were assumed solar abundance.
We made fine tuning of XIS gain offset
and treated cross-normalization of each detector as free parameters.
We ignored the Galactic absorption column to the direction of N132D
\citep[1.6$\times 10^{21}$~xm$^{-2}$;][]{kalberla2005},
since it increases flux only 3\% in the 2--10~keV,
and 0.3\% in the 6--7~keV band, which we are especially interested in
for the Fe K line.
Both models well reproduce the observed spectrum.
The gain offset obtained is 3--5~eV for FI and ~9~eV for BI,
which is roughtly within the range of the XIS gain uncertainty.
We find small residuals around 6.4~keV. 
We attribute this feature to fluorescence emission from neutral Fe,
and thus added a narrow gaussian with fixed center energy of 6.4~keV.
The fitting was improved, with a significance level of this emission component 
of 0.7$\sigma$ for model (a) and 3.8$\sigma$ for model (b).
Finally, we estimate flux (or its upper limit) of nonthermal X-rays 
by adding a power-law component. 
We fixed the photon index to be 2.4, same value as that in the gamma-ray band.
The fitting was improved with
3.3$\sigma$ detection power-law component for model (a)
and 30$\sigma$ detection for model (b).
The best-fit models and parameters are shown in Figure~\ref{fig:wideband}
and Table~\ref{tab:wideband}, respectively.
We also tried the fittings with the free photon index,
but we could not determine the photon index well
($>$1.8 for model (a) and 1.2--3.0 for model (b) for the single parameter errors).
Both models have similar cstats/d.o.f values and leave no large structure in the residuals.
Therefore, we cannot conclude which model better represents the observed X-ray spectrum 
from the statistical point of view. 
In order to determine the parameters of both thermal and non-thermal component,
we need further observations with excellent energy resolution
such as Hitomi \citep{hitomi2017}.

\begin{figure}
\epsscale{0.7}
\plotone{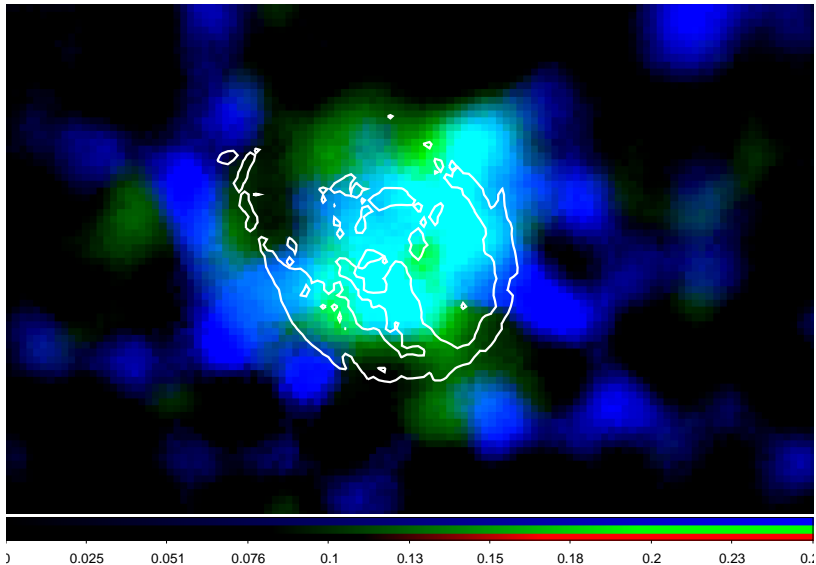}
\caption{%
{\it NuSTAR} images of iron K line (blue) and 10--15~keV (green) bands
in linear scale.
Both images are smoothed with Gaussian model ($\sigma$=9.8~arcsec).
White contour shows the {\it Chandra} image for the comparison.
\label{fig:iron-hardband-image}}
\end{figure}

For the further check of the origin of hard X-rays,
we compared a iron K line image and 10--15~keV band image
of {\it NuSTAR}.
The iron line band was defined as 6.0--7.2~keV
from Figure~\ref{fig:nustarspec}.
The band image also contains the continuum emission.
We estimated the contamination of the continuum component
with the spectral fitting result shown in Table~\ref{tab:nustarlines},
and subtracted.
Figure~\ref{fig:iron-hardband-image} shows
the iron K line (blue) and 10--15~keV band (green) images.
Both show no enhancement on the outer-edge regions
shown in the {\it Chandra} image (white contour).
Further investigation with good statistics and spatial resolution is needed.

\subsection{Comments on pile-up effect on {\it Suzaku} spectra}
N132D is very bright in X-rays, and we should be careful
about pile-up effect especially for the {\it Suzaku} data set,
which can mimic the hard tail in the spectra.
We thus investigate the pile-up effect in {\it Suzaku} spectra
to judge whether the power-law component we detected is
the result of pile-up or not.
As the first step, we have checked the pile-up fraction
with the ftool {\tt pileest},
and found that the fraction is smaller than 4\% for FI
and 6\% for BI.
This is the average value for the all energy band,
whereas our special interest is in harder energy band.
Thus as the next step 
we simulated the piled-up and non-piled up spectra and compared them.
It is found that
the upper-limit of the ratio of flux coming from piled-up events to the total is
$\sim$10\% in 8--10~keV band.
The flux ratio of the power-law component to the total in the 8--10~keV band is 
10\% for model (a) 
and 20\% for model (b).
The former is in the range of expected pile-up,
whereas the latter is still larger than the expected pile-up ratio.
Since we cannot judge which model is better,
We thus treat the larger end of the error region of the power-law flux
as the upper-limit with systematic errors.

\section{Discussion}
\label{sec:discuss}

\subsection{Thermal emission}
\label{sec:thermal}

Thanks to the large effective area of 
{\it NuSTAR},
and low and stable background of {\it Suzaku} XIS above 5~keV,
we detected significant Fe Ly $\alpha$ emission line for the first time.
This suggests the existence of either a very high temperature (model (a)) or
over-ionized (recombining) component (model (b)). Although these models 
cannot be distinguished from statistical point of view, 
the temperature required in model (a) is unusually high as a middle-aged SNR. 
Moreover, this model requires high abundances of Si and S in the low temperature component, 
inconsistent to the previous observations.
We thus assume model (b) to be a reasonable interpretation for the high-temperature 
plasma in N132D and discuss its origin hereafter.

SNRs with recombining plasma have several characteristics.
They are usually associated with GeV gamma-ray emission
coming from molecular clouds.
They also exhibit mixed-morphology (MM) shape
\citep[radio shell plus center-filled thermal X-rays;][]{rho1998}
\citep[c.f.,][]{yamaguchi2009}.
However, N132D does not have a typical MM shape 
but rather a shell-like morphology
in the soft X-ray band \citep{borkowski2007}.
The iron K line image, on the other hand,
shows a center-filled morphology \citep{behar2001,plucinsky2015}.
These facts imply that
the low temperature component mainly studied in previous works
makes the shell-like structure,
whereas the recombining component remains in the center of the remnant.
These results may indicate again that
N132D is in the transition stage from shell-like young SNRs
to old ones with MM morphology;
during the transition, the center-filled emission appears first,
and after that the low temperature emission from the shell
will disappears to make the MM morphology.
In summary, it is expected that N132D is in transition phase.

In order to derive the density and thermal energy,
the uniform density sphere with the radius of 30~arcsec or 7~pc at 48~kpc was assumed \citep{macri2006}.
We derived the density from the emission measure $\sim$3~cm$^{-2}$.
The thermal electron energy of the recombining component,
$3 n_e kT$ times volume,
is estimated to be $\sim 9\times 10^{50}$~erg.
Combining those of lower temperature components
\citep{behar2001,borkowski2007},
the total thermal energy exceed $\sim 10^{51}$~erg.
Moreover, the kinetic energy should be dominant
when SNRs are young,
thus the total energy exceeds
$10^{51}$~erg even more.
This result implies that the progenitor explosion of N132D
might be more energetic than usual supernovae.

\subsection{Nonthermal emission}
\label{sec:nonthermal}

N132D is the brightest SNR in the gamma-ray band
\citep{fermi2015,hess2015}.
We estimated an upper-limit on the nonthermal X-ray flux
of $7.3\times 10^{-13}$~ergs~s$^{-1}$cm$^{-2}$
or $2.0\times 10^{35}$~erg~s$^{-1}$ in the 2--10~keV band.
This is fainter than the younger samples, Cas~A
\citep{helder2008}.
\citet{nakamura2012} suggests that
the nonthermal X-ray luminosity becomes fainter
when SNRs get older,
thus our faint nonthermal X-rays in N132D is
consistent with the older age.
The primary nonthermal emission mechanism in the X-ray band 
is synchrotron from accelerated electrons,
whereas GeV and VHE gamma-rays have leptonic origin via inverse Compton and/or hadronic origin
via $\pi^0$ decay, respectively.
The ratio between TeV gamma-ray flux in 1--10~TeV ($F_{TeV}$)
and synchrotron X-ray flux in 2--10~keV
($F_X$), $F_{TeV}/F_X$, is useful
to resolve these two possibilities \citep{yamazaki2006,matsumoto2007,bamba2007}.
In our case, $F_{TeV}/F_X$ is larger than 7,
which is much larger than
young SNRs with strong synchrotron X-ray filaments
\citep{yamazaki2006}.
This result implies that
gamma-rays from N132D have a hadronic origin.
This agrees with the observational fact that
the GeV gamma-ray luminosity is $\sim 10^{36}$~erg~s$^{-1}$,
which is very difficult to reproduce with the inverse compton scattering
of the cosmic microwave background
\citep[$\sim 10^{35}$~erg~s$^{-1}$; ][]{acero2016,bamba2016}.
A more detailed discussion will be found in the following.

\begin{figure}
\epsscale{0.5}
\includegraphics[width=0.5\textwidth]{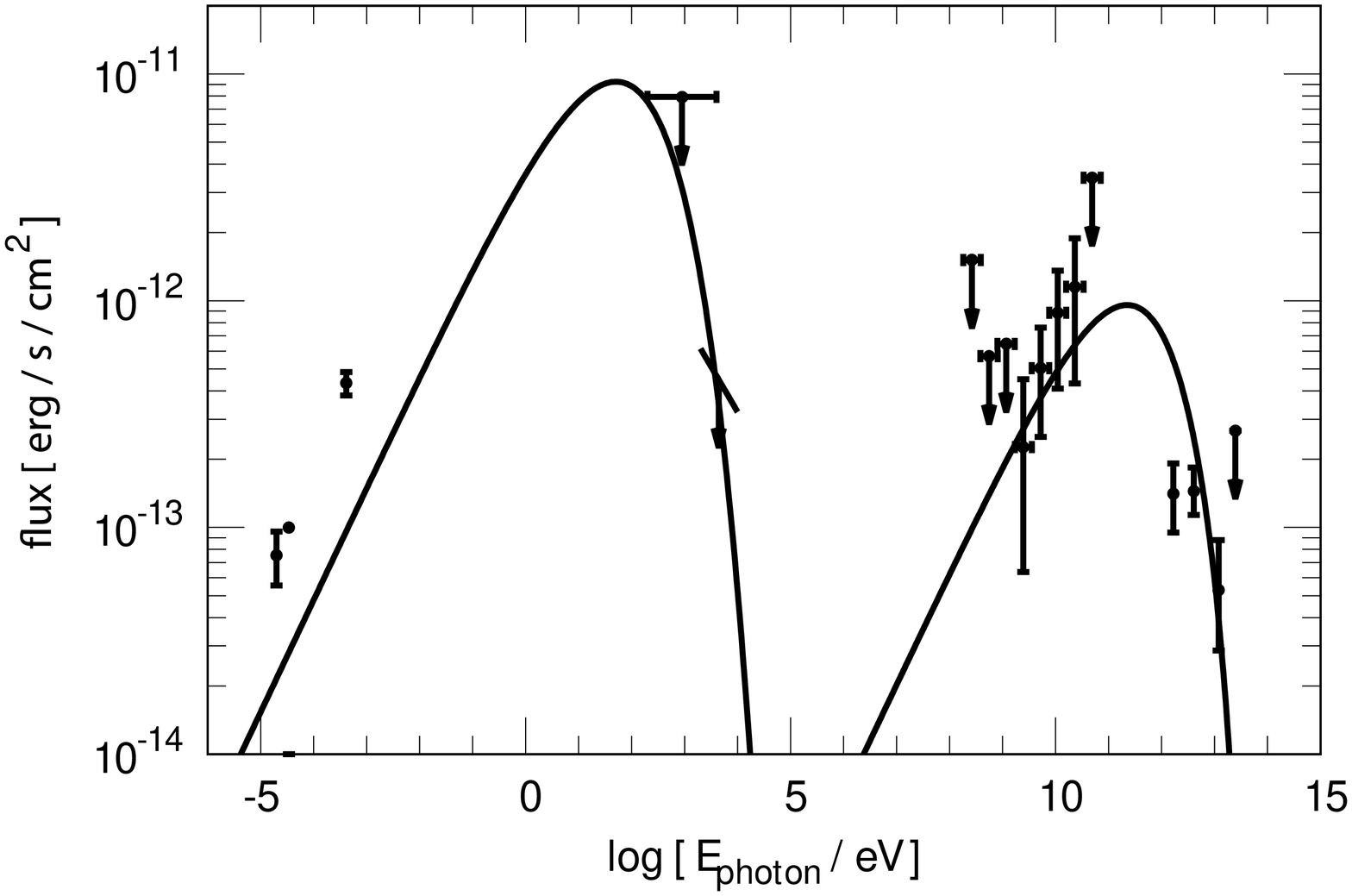}
\includegraphics[width=0.5\textwidth]{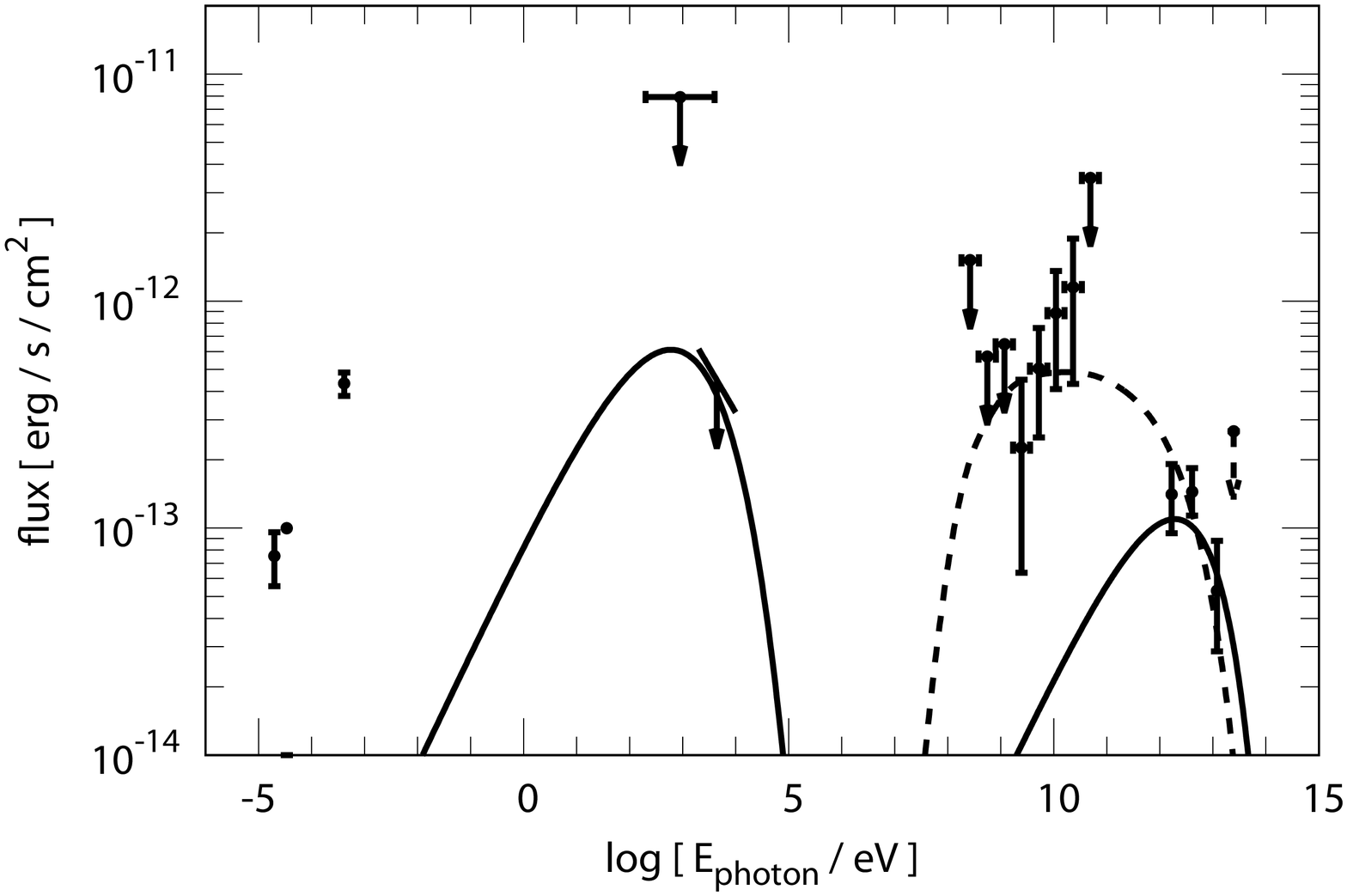}
\caption{%
SED of N132D with pure leptonic (left) and leptonic + hadronic models (right),
respectively
\citep{dickel1995,planck2014,hughes1998,fermi2015,hess2015}.
Solid and dashed lines represent
leptonic and hadronic components, respectively.
In the left panel,
solid lines are for $U_e$ of $2.5\times 10^{49}$~erg,
cut-off energy of electrons of 7~TeV,
and the magnetic field of 20~$\mu$G,
respectively.
In the right panel,
solid lines are for
$U_e$ of $10^{48}$~erg, the magnetic field of 13~$\mu$G,
and the maximum electron energy of 30~TeV,
whereas dashed lines are for
the maximum proton energy of 30~TeV, the total proton energy of $10^{50}$~erg,
and the density of surrounding matter of 80~cm$^{-3}$.
Although the pure leptonic model
on the left shows good agreement with the data,
it is physically unrealistic because
it requires too much energy in the electrons.
\label{fig:sed}}
\end{figure}

Figure~\ref{fig:sed} shows
the broad-band spectral energy distribution (SED) of N132D
for a more quantitative study.
3.5~cm and 6~cm data were taken from \citet{dickel1995},
whereas we used {\it PLANCK} data points at 100~GHz \citep{planck2014}.
\citet{planck2014} used 2.6~arcmin radius region for the source spectra,
which includes surrounding molecular clouds.
We thus used this data point as an upper-limit.
Our analysis concentrates on the hard X-rays,
whereas {\it ASCA} shows the upper-limit of nonthermal emission
in the 0.5--5~keV band with the assumption of $\Gamma=2$ \citep{hughes1998}.
{\it Fermi} and H.E.S.S. data points were also used \citep{fermi2015,hess2015}.

The solid lines in the left panel of Figure~\ref{fig:sed} represent a pure leptonic model to reproduce the data points.
We assumed the photon field of inverse compton emission
in the same way of \citet{hess2015},
the infrared emission from dust in and around the SNR.
This model requires the magnetic field of $\le$20~$\mu$G,
the cut-off energy for electron of 7~TeV,
and the total amount of energy of electrons ($U_e$) of $2.5\times 10^{49}$~erg.
The derived $U_e$ is very large
and unrealistic in a typical SNR,
since we should also consider the energy of accelerated protons,
which should be comparable or larger than that of electrons
\citep{slane2016}.
The total electron energy does not change
even if we assume different spatial distribution
for electrons and protons.
These results suggest that the emission cannot have a pure leptonic origin, but instead must have a hadronic one.
This result is mainly from bright GeV gamma-rays
and faint synchrotron X-rays,
which require a small magnetic field and as a result large $U_e$.

The right panel of Figure~\ref{fig:sed} represents
the leptonic + hadronic model fit to the gamma-ray data.
We estimated the maximum energy of accelerated proton $E_{max}$ first from GeV--VHE gamma-ray spectrum,
and determined the magnetic field and the amount of electrons
with the assumption that electrons and protons have the same maximum energy.
We then estimated the inverse compton emission
from accelerated electrons and the same infrared photon field
to that in our pure leptonic model analysis \citep{hess2015}.
When we fix the proton spectral index to be 2.0
and a total proton energy of $10^{50}$~erg,
we estimated
the number density of 80~cm$^{-3}$ for the pre-shock interstellar gas
and $E_{max}$ of 30~TeV.
Assuming electron cut-off energy is same to $E_{max}$,
we derived $U_e$ of $\le 10^{48}$~erg and  magnetic field of $\le 13~\mu$G.

The presence of a molecular cloud
around the SNR supports the large number density of ambient matter.
The best-fit proton energy is $\sim$10\% of
the typical explosion energy.
Note that the derived energy is only for those emitting gamma-rays,
so that more energy is required if there are accelerated protons 
which do not emit gamma-rays.
These facts suggest that
N132D is a very energetic cosmic ray accelerator.
This result can be connected to our results from thermal aspects,
that the progenitor explosion was more energetic than
conventional supernovae.

\begin{figure}
\epsscale{0.7}
\plotone{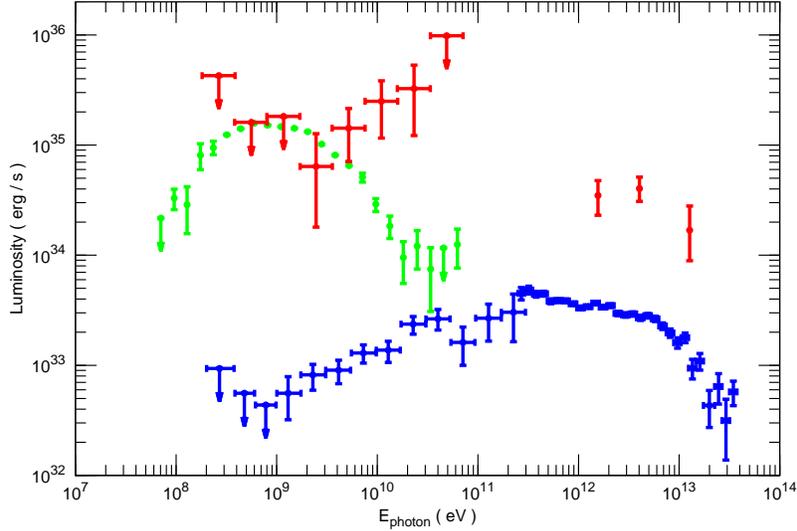}
\vspace{10mm}
\caption{%
Gamma-ray spectra of young SNR sample (RX~J1713$-$3946; blue),
middle-aged sample (W44; green),
and N132D (red).
The horizontal axis represents 4$\pi d^2 \times \nu F_\nu$,
where $d$, $\nu$, and $F_\nu$ are
distance, frequency, and differential flux.
\label{fig:sed_hikaku}}
\end{figure}

Many old SNRs with gamma-rays are now identified
as emitting from accelerated protons with the maximum energy of $\sim$10~GeV
(for example, W44 \citep{abdo2010b}, W28 \citep{abdo2010c}.
On the other hand,
the gamma-ray spectrum of N132D 
is harder than the spectra from such old SNRs,
extending up to $\sim$ TeV range.
Figure~\ref{fig:sed_hikaku} shows the SED of RX~J1713$-$3946 as typical young SNR case \citep{hess2016},
W44 as a typical middle-aged \citep{ackermann2013},
and N132D.
It is widely agreed that the gamma-ray emission from W44 is hadronic origin,
whereas
in the case of RX J1713-3946, there is still debate about whether it is leptonic \citep{ellison2010,ohira2017} or hadronic \citep{inoue2012,gabici2014,yamazaki2009}.
One can see that the cut-off energy of the gamma-ray spectrum of N132D
is between those of W44 and RX~J1713$-$3946.
Actually, the maximum energy of accelerated protons in N132D (30~TeV)
is between that for RX~J1713.7$-$3946 \citep[$>$100~TeV;][]{aharonian2007} and
W44 \citep[7--9~GeV;][]{abdo2010b}.
Our comparison implies that
N132D is in a transition phase from young to old,
also in the context of particle acceleration.

Protons are accelerated to high energies
when SNRs are young, and start escaping
when SNRs get older or interact with molecular clouds \citep{ptuskin2005,ohira2010,ohira2012}
and emit gamma-rays \citep{gabici2009,ohira2011}.
Since N132D is in the transition phase,
the remnant has already accelerated protons to very high energies,
and still produces them at a rate higher than it loses them.
This may explain the exceptionally bright gamma-ray emission from N132D.

\subsection{Comments on the neutral iron K line}

We detected possible signal of neutral iron K line from N132D.
The luminosity of the line is $\sim 1.9\times 10^{33}$~erg~s$^{-1}$
assuming a distance of 48~kpc \citep{macri2006}.

Neutral iron K line from SNRs is rather peculiar,
since the SNR plasma is highly ionized.
RCW~86 has strong neutral iron K emission \citep{bamba2000},
which is from plasma with very low ionization time scale
\citep{yamaguchi2008}.
The line luminosity is $3.2\times 10^{32}$~erg~s$^{-1}$ \citep{ueno2007}
with the distance of 2.3~kpc \citep{sollerman2003},
which is similar to that in our case.
On the other hand,
this low ionization plasma in RCW~86 has been heated very recently
via the interaction with the cavity wall \citep{broersen2014},
which is not our case since the main plasma in N132D is over-ionized.

\begin{figure}
\epsscale{0.7}
\plotone{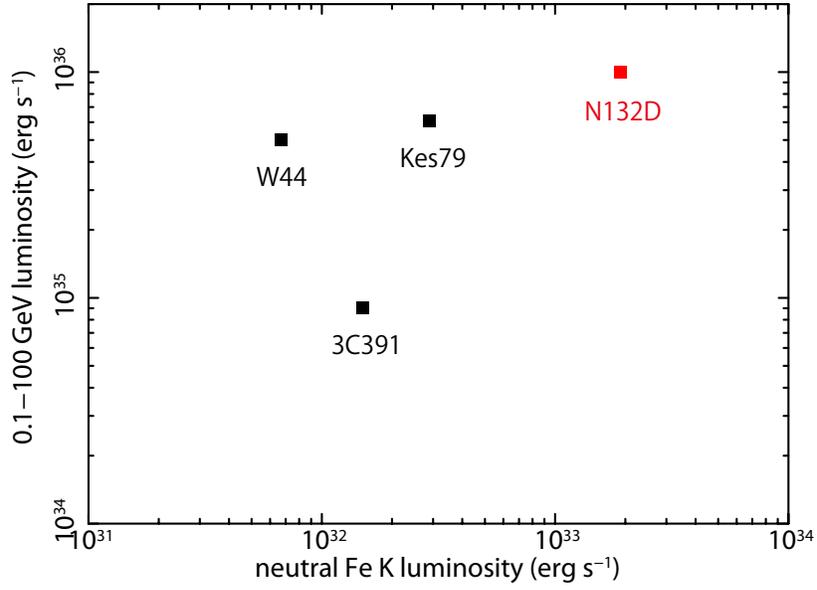}
\caption{%
Relation between the neutral iron K line luminosity and
0.1--100~GeV gamma-ray luminosity.
Red mark represents N132D \citep[this work, ][]{fermi2015},
whereas black ones are for Kes~79, W44, and 3C391 \citep{sato2016a,sato2016,auchettl2014,acero2016}.
\label{fig:iron-gev}}
\end{figure}

Other samples with the neutral iron K line are
3C391 \citep{sato2014} and Kes~79 \citep{sato2016}.
The former has the line luminosity of
$\sim 1.5\times 10^{32}$~erg~s$^{-1}$ \citep{sato2016a} at 8~kpc
\citep{reynolds1993},
whereas the latter has the line luminosity of
$2.8\times 10^{32}$~erg~s$^{-1}$ at 7.5~kpc
\citep{giacani2009}, slightly smaller
but comparable within the large error region of our case.
\citet{sato2016} claims the possibility that
the line from Kes~79 is
due to K-shell ionization of neutral iron
by the interaction of low-energy cosmic-ray protons
with the surrounding molecular cloud,
since the peak position of the line emission coincides with
the molecular cloud \citep{giacani2009},
together with the detection of GeV gamma-rays \citep{auchettl2014}.
Similar discussion can be done for 3C391,
together with OH masers \citep{frail1996} and GeV emission
\citep{castro2010}.
W44 is also similar to this sample with the possible detection of
neutral iron K line
\citep[$7.0\pm5.0\times 10^{-6}$~ph~cm$^{-2}$s$^{-1}$;][]{sato2016a}
and GeV gamma-ray emission \citep{acero2016}.
If these lines are due to low-energy cosmic-ray protons, 
the luminosities of neutral iron K line and GeV gamma-ray
may have a positive correlation,
since the former is roughly propotional to the number of 
low-energy cosmic-ray protons
in the MeV range and target mass
\citep[e.g.,][]{valinia2000,dogiel2011,nobukawa2015}
whereas the latter to the number of GeV protons and target mass.
Figure~\ref{fig:iron-gev} shows the relation between
neutral iron K line and 0.1--100~GeV gamma-ray luminosities
of these SNRs.
Although the sample number is too small and the error is large,
the tendency is consistent with our scenario.
It can be the first clue of the proton acceleration in the MeV range in SNRs.
In order to study this issue further,
we need more X-ray observations with good energy and spatial resolution
and better statistics.

\section{Conclusions}
We have performed X-ray spectroscopy 
of the SNR N132D in the LMC,
using {\it NuSTAR} and {\it Suzaku}
deep observations.
Thanks to the large effective area
and low, stable background,
we discover that a very high temperature plasma or recombining plasma
with $nt$ $\sim$8.8$\times 10^{11}$~cm$^{-3}$s.
Together with the morphology of the iron K emission,
we coclude that
N132D is in the transition phase
from young shell-like SNR with ionizing plasma
to middle-aged mixed morphology SNR with recombining plasma.
The total thermal energy of $> 10^{51}$~erg shows
the progenitor explosion of N132D can be very energetic.
We have made the tight upper-limit of nonthermal X-rays.
The spectral energy distribution from radio to VHE gamma-rays shows us that
the total energy of accelerated protons is very energetic,
implying that N132D is a very efficient proton accelerator.
The cut-off energy in the gamma-ray band also shows that
N132D is in a transition stage from young to middle-aged.

\acknowledgments

We thank the anonymous referee for the fruitful comments.
We thank The TeGeV Catalogue at ASDC (v2)
(http://tools.asdc.asi.it/catalogSearch.jsp) to make the SED.
This research has made use of
NASA's Astrophysics Data System Bibliographic Services,
and the SIMBAD database,
operated at CDS, Strasbourg, France.
This work is supported in part by
Grant-in-Aid for Scientific Research of
the Japanese Ministry of Education, Culture, Sports,
Science and Technology (MEXT) of Japan,
No.~22684012 and 15K051017 (A.~B.), 
16K17702 (Y.~O.), 15K05088 (R.~Y.),
15K17657 (M.~S.), and 16K17673 (S.~K.).
E.~M. acknowledges support from NASA grant NNX15AC76G.


\begin{deluxetable}{lcccc}
\tabletypesize{\scriptsize}
\tablecaption{Observation log\label{tab:obs}}
\tablewidth{0pt}
\tablehead{
\colhead{Satellite} & \colhead{ObsID} & \colhead{Date} & \colhead{Position} & \colhead{Exposure}\\
& & (YYYY/MM/DD) & (Ra, Dec.) & (ks)
}
\startdata
{\it NuSTAR} & 4010101000 & 2015/12/10 & (81.3115, $-$69.6662) & 62.3 \\
{\it Suzaku} & 105011010 & 2010/07/27 & (81.2790, $-$69.6505) & 35.8 \\
{\it Suzaku} & 106010010 & 2011/04/25 & (81.2700, $-$69.6453) & 26.0 \\
{\it Suzaku} & 106010020 & 2011/10/07 & (81.2448, $-$69.6458) & 23.9 \\
{\it Suzaku} & 107008010 & 2012/10/19 & (81.2348, $-$69.6540) & 24.1 \\
{\it Suzaku} & 107008020 & 2013/03/26 & (81.2721, $-$69.6340) & 23.1 \\
{\it Suzaku} & 108008020 & 2013/10/06 & (81.2389, $-$69.6575) & 5.1 \\
{\it Suzaku} & 108008030 & 2013/05/22 & (81.2957, $-$69.6416) & 29.8 \\
{\it Suzaku} & 108008040 & 2013/11/25 & (81.2290, $-$69.6452) & 27.1 \\
{\it Suzaku} & 108008050 & 2014/01/11 & (81.2306, $-$69.6358) & 7.5 \\
{\it Suzaku} & 109009010 & 2014/04/17 & (81.2854, $-$69.6352) & 24.9 \\
{\it Suzaku} & 109009020 & 2014/10/30 & (81.2260, $-$69.6524) & 13.5
\enddata
\end{deluxetable}

\begin{deluxetable}{p{6pc}c}
\tabletypesize{\scriptsize}
\tablecaption{Best-fit parameters for the NuSTAR spectra\tablenotemark{a}
\label{tab:nustarlines}}
\tablewidth{0pt}
\tablehead{
\colhead{Parameters} 
}
\startdata
Bremss\\
\hspace{3mm}$kT$ (keV)\dotfill & 1.71 (1.65--1.78) \\
\hspace{3mm}Norm\tablenotemark{b}\dotfill & 7.6 (6.9--8.3) \\
Gaussian 1\\
\hspace{3mm}$E_c$ (keV)\dotfill & 6.655 (6.641--6.677) \\
\hspace{3mm}Flux\tablenotemark{c}\dotfill & 13.4 (11.9--14.9) \\
Gaussian 2\\
\hspace{3mm}$E_c$ (keV)\dotfill & 7.92 (7.81--7.98) \\
\hspace{3mm}Flux\tablenotemark{c} & 2.3 (1.5--3.2) \\
cstat/d.o.f.\dotfill & 287.8/131
\enddata
\tablenotetext{a}{Errors indicate single parameter 90\% confidence regions.}
\tablenotetext{b}{Emission measure in the unit of
$\frac{3.02\times 10^{-15}}{4\pi D^2}\int n_en_idV$,
where $D$ is the distance to the source (cm)
and $n_e$, $n_i$ are the electron and ion densities (cm$^{-3}$).}
\tablenotetext{c}{In the unit of $10^{-6} $photons~cm$^{-2}$s$^{-1}$.}
\end{deluxetable}

\begin{deluxetable}{p{6pc}c}
\tabletypesize{\scriptsize}
\tablecaption{Best-fit parameters for the Suzaku spectra\tablenotemark{a}
\label{tab:suzakulines}}
\tablewidth{0pt}
\tablehead{
\colhead{Parameters} 
}
\startdata
Bremss\\
\hspace{3mm}$kT$ (keV)\dotfill & 2.25 (2.02--2.35) \\
\hspace{3mm}Norm\tablenotemark{b}\dotfill & 3.4 (3.0--4.3) \\
Gaussian 1\\
\hspace{3mm} $E_c$ (keV)\dotfill & 6.652 (6.648--6.660) \\
\hspace{3mm} Flux\tablenotemark{c}\dotfill & 14.8 (14.0--15.5) \\
Gaussian 2\\
\hspace{3mm} $E_c$ (keV)\dotfill & 6.92 (6.88--6.98) \\
\hspace{3mm} Flux\tablenotemark{c}\dotfill & 1.2 (0.8--1.7) \\
Gaussian 3\\
\hspace{3mm} $E_c$ (keV)\dotfill & 7.85 (7.77--7.93) \\
\hspace{3mm} Flux\tablenotemark{c}\dotfill & 0.9 (0.5--1.4) \\
cstat/d.o.f.\dotfill & 145.4/109
\enddata
\tablenotetext{a}{Errors indicate single parameter 90\% confidence regions.}
\tablenotetext{b}{Emission measure in the unit of
$\frac{3.02\times 10^{-15}}{4\pi D^2}\int n_en_idV$,
where $D$ is the distance to the source (cm)
and $n_e$, $n_i$ are the electron and ion densities (cm$^{-3}$).}
\tablenotetext{c}{In the unit of $10^{-6}$photons~cm$^{-2}$s$^{-1}$.}
\end{deluxetable}

\begin{deluxetable}{p{12pc}cc}
\tabletypesize{\scriptsize}
\tablecaption{Best-fit parameters for the combined {\it Suzaku} and {\it NuSTAR} spectra\tablenotemark{a}
\label{tab:wideband}}
\tablewidth{0pt}
\tablehead{
\colhead{Parameters} & \colhead{model (a)\tablenotemark{b}} & \colhead{model (b)\tablenotemark{b}}
}
\startdata
{low $kT$ comp.}\\
\hspace{3mm} $kT$ (keV)\dotfill & 0.54 (0.39--0.56) & 0.70 (0.65--0.71) \\
\hspace{3mm} $Z_{\rm Si}$\dotfill & 1.9 ($>$1.4) & 0.63 (0.59--0.74)  \\
\hspace{3mm} $Z_{\rm S}$\dotfill & 2.1 ($>$1.8) & 0.91 (0.86--1.10) \\
\hspace{3mm} Norm\tablenotemark{c}\dotfill & 3.3 (0.1--4.1) & 5.6 (4.7--5.8) \\
{middle $kT$ comp.}\\
\hspace{3mm} $kT$ (keV)\dotfill & 1.2 (0.8--1.2) & 1.5 (1.3--1.6) \\
\hspace{3mm} $kT_{init}$ (keV)\dotfill & --- & $>$ 8 \\
\hspace{3mm} $Z_{\rm Si}=Z_{\rm S}$\dotfill & 0.45 (0.40--0.48) & 0.42 (0.36--0.49) \\
\hspace{3mm} $Z_{\rm Fe}=Z_{\rm Ni}$\dotfill & 0.74 (0.61--0.96) & 0.46 (0.40--0.57) \\
\hspace{3mm} $n_{\rm e}t$ ($10^{11}$s~cm$^{-3}$)\dotfill & --- & 8.8 (7.0--10.0) \\
\hspace{3mm} Norm\tablenotemark{c}\dotfill & 2.8 (2.5--4.0) & 1.3 (1.2--1.7) \\
high $kT$ comp.\\
\hspace{3mm} $kT$ (keV) \dotfill & 5.7 (4.0--6.8) & --- \\
\hspace{3mm} Norm\tablenotemark{c}\dotfill & 0.06 (0.05--0.10) & --- \\
Neutral iron line\\
\hspace{3mm} Flux\tablenotemark{d}\dotfill & 1.5 ($<$5.7) & 6.7 (2.0--11.0) \\
Power-law\\
\hspace{3mm} $\Gamma$\dotfill & 2.4 (fixed) & 2.4 (fixed) \\
\hspace{3mm} $F_{\rm 2-10keV}$ ($10^{-13}$~erg~cm$^{-2}$s$^{-1}$)\dotfill & 1.9 (0.5--4.1) & 5.0 (4.2--7.3) \\
Gain fit for XIS FI (eV)\dotfill & 3.9 & 5.4 \\
Gain fit for XIS BI (eV)\dotfill & 11.7 & 9.8 \\
cstat/d.o.f.\dotfill & 419.5/323 & 431.3/323
\enddata
\tablenotetext{a}{Errors indicate single parameter 90\% confidence regions.}
\tablenotetext{b}{Model (a) is 3 {\tt vapec} model, and
Model (b) is 2~{\tt apec} + {\tt vrnei} + power-law model.}
\tablenotetext{c}{Emission measure in the unit of
$\frac{10^{-16}}{4\pi D^2}\int n_en_idV$,
where $D$ is the distance to the source (cm)
and $n_e$, $n_i$ are the electron and ion densities (cm$^{-3}$).}
\tablenotetext{d}{In the unit of $10^{-7}$photons~cm$^{-2}$s$^{-1}$.}
\end{deluxetable}

\end{document}